\documentstyle[floats,aps,prl,twocolumn,epsfig]{revtex} 
\draft
\begin{document}
%\tighten
\title{Observation of emission from chaotic lasing modes in deformed 
microspheres: displacement by the stable orbit modes}
\author{Seongsik Chang$^{\ast}$, 
Jens U. N\"{o}ckel$^{\dag}$, Richard K. Chang$^{\ast}$, 
and A. Douglas Stone$^{\ast}$}
\address{$^{\ast}$ 
Department of Applied Physics, Yale University, New Haven, 
Connecticut 06520\\
$^{\dag}$
Max-Planck-Institut f\"{u}r Physik komplexer Systeme,Dresden, 
Germany D-01187
}
\date{\today}
\maketitle

\begin{abstract}
By combining detailed imaging measurements at different 
tilt angles with simulations of ray emission from prolate deformed 
lasing micro-droplets, we conclude that the probability density 
for the lasing modes in a three-dimensional dielectric microcavity must 
reside in the chaotic region of the ray phase space.  In particular, 
maximum emission from such chaotic lasing modes is not from tangent 
rays emerging from the highest curvature part of the rim.   The 
laser emission is observed and calculated to be non-tangent 
and displaced from the highest curvature due to the 
presence of stable orbits. In this Letter we present the first 
experimental evidence for this phenomenon of ``dynamical eclipsing''.
\end{abstract}


Both the Schr{\"o}dinger equation and the wave equation 
of optics share the property that their short wavelength limit 
corresponds to a Hamiltonian dynamics: classical Newtonian mechanics 
in the former case and ray-optics in the latter. When that dynamics 
is non-integrable, relating  properties of the wave 
solutions to the classical dynamics is very difficult and has 
been an area of intense study under the rubric {\it quantum/wave chaos}
\cite{mcg}.
Of particular interest recently is the generic 
non-integrable case of a mixed phase space in which chaotic and 
regular regions coexist.  It was 
noted\cite{optlet1,prl,optlet2,optchap,nature}
by several of the authors that an interesting problem 
in optics, the resonance and emission properties of deformed 
spherical and cylindrical dielectric resonators, was just such 
a problem in wave chaos theory with a mixed phase space, with 
resonant modes corresponding to both regular and chaotic regions 
of phase space.   It was predicted \cite{prl,optchap,nature}
that in a substantial parameter range deformed micro-cylinder 
and micro-spherical lasers would actually lase on a certain type 
of chaotic mode which we refer to as ``chaotic whispering gallery 
modes'' (CWGMs).   In this Letter, we report for the first time 
laser emission patterns unique to the CWGM of deformed 
micro-droplet lasers.   The laser emission from CWGMs is found 
to be non-tangent and its emission location is displaced from 
the points of highest curvature where most of the laser emission 
from regular whispering gallery modes occurs. 

 Whispering gallery modes (WGMs) are solutions of 
the wave equation which classically correspond to rays circulating 
around the boundary reflecting at high angle of incidence; in 
the dielectric resonator such modes have long lifetimes due to 
quasi-total internal reflection at the curved boundary.   If angular 
momentum is conserved, then a metastable well is formed by the 
centrifugal barrier and the dielectric/air interface; quasi-bound 
states within this well decay only by tunneling (evanescent leakage)
\cite{johnson}.
If the dielectric is deformed, angular momentum 
is no longer conserved and such a state may emit ``over the barrier`` 
by refraction, a ``classically allowed`` process.   Light emission 
via refraction is non-tangent to the interface, 
whereas that via tunneling is always tangential. 
Our observation of spatially displaced non-tangent laser 
emission is a signature 
of new  types of lasing modes in this system unique to 
the deformed case. 

A CWGM is a solution of the wave equation corresponding 
to an ensemble of rays in a chaotic region of phase space; and 
the lifetime of such modes is, for large enough deformation, limited 
by refractive escape\cite{nature}.  
For a deformed cylindrical resonator, such a 
mode obtained by exact numerical solution of the wave equation, 
is depicted in Fig. 1(a).  Superimposed on a surface of section 
(SOS) of the phase space for ray motion, we show the Husimi function 
of the mode\cite{leboef},
which is a projection of the exact wave solution 
for a resonant mode onto this SOS.  As can be seen in Fig. 1(a), 
the mode is localized away from the critical angle for total 
internal reflection, $\chi_{c}$, but its tail leaks downwards to 
$\chi_{c}$ in a highly anisotropic fashion, leading to directional 
emission\cite{optlet2,nature} 
[see the inset of Fig 1(a)]. Previous work analyzed 
anisotropic emission patterns from deformed lasing droplets\cite{prl} 
and dye jets\cite{optlet2} and found results consistent with, but not unique 
to, lasing from chaotic modes. Rather recently, deformed cylindrical 
semiconductor lasers\cite{gmachl}
were found to exhibit highly directional 
and high-power output, but feedback was provided by a mode associated 
with a non-chaotic ``bow-tie`` orbit.

\begin{figure}[bt]
\epsfig{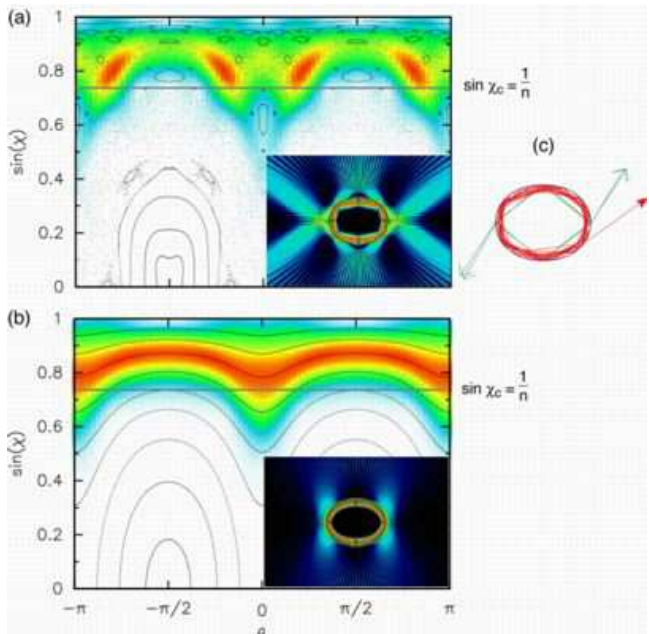}
\caption{
Husimi plots (red: high intensity, white: no weight) 
superimposed onto classical surfaces of section for the 
ray dynamics inside 2D deformed dielectric cavities with refractive
index $n=1.36$. (a) CWGM at $ka = 52.85$
in a quadrupolar cavity parametrized by $r(\theta) = 1 + \epsilon\cos
2\theta$ with $\epsilon = 0.113$. (b) WGM at $ka=52.04$ in an ellipse 
with $b/a = 1.27$. 
Insets: the corresponding wave functions in real space. (c) 
Sketch of quasi-periodic four-bounce orbits (green) and a chaotic orbit
(red) inside the quadrupolar cavity of (a).
}
\end{figure}
The main problem with interpreting observations 
from microlasers with deformed resonators is that highly directional 
emission may also be obtained with non-chaotic ray dynamics such 
as an elliptically-deformed cylinder [see Fig. 1(b)].  It is 
thus desirable to find a distinct signature of lasing from chaotic 
modes.  One such signature\cite{optlet2,nature} is illustrated 
by the SOS in Figs. 1(a) and 1(b).  In the SOS of Fig. 1(a), 
the islands which represent orbits near stable periodic paths 
prevent the entry of ray flux from the chaotic regions 
(neglecting weak tunneling effects).  
This effect can {\it displace } 
the rays of such a chaotic mode from the points 
of highest curvature at which generically high emission would 
be expected to occur. We refer to the scenario of Fig.1(a) as 
{\it dynamical eclipsing}.  This displacement of the chaotic mode rays also 
has a dramatic effect on the laser emission pattern, which can 
be determined experimentally by taking images of the microcavity 
at various angles. 

The identical ray dynamics underlying Fig. 1(a) 
for a two-dimensional (2D) microstructure also occurs in a plane 
that cuts an axisymmetric quadrupole-deformed droplet from pole 
to pole through a meridian (polar plane). Such droplets made 
of ethanol (refractive index, $n= 1.36$) containing laser dye are 
ejected from the 
vibrating orifice of a Berglund-Liu droplet generator\cite{berglund}, 
oscillating in shape\cite{prl}.  The dominant multipole components of
this hydrodynamic oscillation\cite{lamb} are quadrupolar and octupolar
with 
damping time constants $\tau= 213\,\mu$sec for the quadrupole and
$\tau = 39\,\mu$sec for the octupole.  
For times $t > 100\,\mu$sec during which the experiments 
were performed, 
the shape deformation is essentially purely quadrupolar, and 
is uniquely parameterized by the aspect ratio, $b/a$, where $b$ 
and $a$ are the semimajor and -minor axes. A particular $b/a$
can be selected by observing the droplets stroboscopically 
at the proper phase of their shape oscillation.

To detect dynamical eclipsing in the polar plane 
of such deformed droplets it is necessary to measure both emission 
directions and locations on the spheroid. The lasing microdroplets 
are viewed at various polar angles ($\theta_D$) 
by using a CCD camera.   2D imaging is essential 
to determine the emission locations, which cannot be provided 
by a single-photodiode experiment.  The experimental setup is 
shown in Fig. 2 where the polar plane is defined to be the Y-Z 
plane.   Variation of the polar angle $\theta_{D}$
of the imaging direction is achieved by fixing 
the camera, but tilting the droplet generator about the pump 
laser direction (X-axis), which preserves the same (non-uniform) 
pumping condition at all tilt angles. Simultaneous images were  
taken in two directions: (I-1) along the X-axis and (I-2) along the 
(-Y)-axis, which were then brought side by side with a beam splitter 
(BS) to the single CCD camera. While the shape and the tilt of 
the droplets were monitored by (I-1), the desired lasing images 
at various values of $\theta_{D}$ are obtained from (I-2).
\begin{figure}[bt]
\epsfig{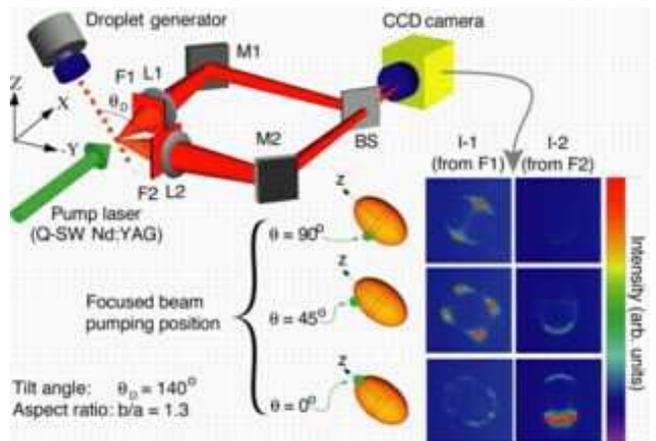}
\caption{A schematic of experimental setup for taking images of lasing
micro-droplets at various inclination angles ($\theta_D$). 
(L1, L2: camera lenses; F1, F2: color filters; 
M1, M2: mirrors; BS: beam splitter). Color filters
were used to block scattered pump laser light. Focused pumping with 
$\!< 5\,\mu$m beam diameter was used for the experiment. Images (I-1
and I-2) of prolate droplet ($b/a = 1.3$) at $\theta_D = 140^{\circ}$ 
are shown
for three different locations of the focused pump beam; (i) equatorial rim,
(ii) $45^{\circ}$ above the equator, and (iii) north pole.}
\end{figure}

We only present data for prolate droplets, for which 
the longest axis goes through the poles; similar effects are 
seen in oblate droplets but are not presented here due to space 
limitations. Taking the long axis as the z-axis, rotational symmetry 
implies that ray orbits can be classified by the conserved 
z-component of the {\it angular momentum}, $L_{z}$\cite{prl}. The 
{\it near-polar orbits} of interest correspond to $L_{z}\approx 0$,
 since they correspond to motion in a plane 
passing near or through the z-axis.  
Experimentally, lasing in 
a particular $L_{z}$-mode can be excited by aiming a tightly focused 
pump laser beam along the droplet rim as shown in Fig 2, which 
subsequently enters the droplet with $\chi\approx\chi_c$ 
and fixes $L_{z}$ to a value of $r\,\sin\theta\sin\chi_c$, where
$r(\theta)$ is the distance from the droplet center to the surface 
($L_{z}$ in these units is equal to the distance of closest 
approach of the rays to the z axis\cite{prl}).   Because $L_{z}$ 
of such a pump beam is conserved after entering 
the droplet, this input beam generates the highest gain for lasing 
modes with the same $L_{z}$.   For example (see Fig. 2), when pumped at the 
equatorial rim, lasing occurs in 
modes with 
$L_{z}\approx a\,\sin\chi_c$
  in which rays are confined to the equatorial plane; 
when pumped tangentially near $\theta= 45^{\circ}$,
lasing occurs in $L_{z}\approx 0.5\,a$ modes for which rays 
precess\cite{swindal}
around the z-axis due to non-zero torque exerted 
on the rays at each bounce.   In the the $L_{z}=0.5\,a$ case, 
the lasing mode is still regular and 
rays encounter highest curvature near the highest ($\theta =
45^\circ$) 
or lowest ($\theta = 135^\circ$) latitudes of the orbits and thus 
escape tangentially through tunneling at these points\cite{ss2}.

In order to demonstrate dynamical eclipsing, near-polar 
modes ($L_{z}\approx 0$) are excited with a focused pump beam aimed at 
the north pole.   In this case the ray motion is that of an effective 
2D cavity of aspect ratio $b/a\approx 1.3$ formed by the intersection 
of the droplet with the polar plane, and is identical to the 
parameters used in Fig. 1(a). No laser emission is observed in 
I-1, but intense laser emission is observed from the bottom of 
I-2 (see Fig.2).  Such an emission pattern indicates that main 
lasing emission does not occur from the highest curvature points 
(north pole or south pole), but away from these points, consistent 
with the prediction of dynamical eclipsing in Fig 1(a).  To explain 
this in more detail, note first that there exists a four-bounce 
stable diamond-shaped orbit for rays in the polar plane 
[Fig. 1(c)], which can support stable resonant modes; however 
these modes do not provide sufficient feedback for 
lasing because of the large refractive leakage near the pole. 
Specifically, the round-trip loss for the four-bounce orbit 
is $4.6$, which 
is significantly greater than the estimated maximum round-trip gain of
$0.9$ for $5\times 10^{-4}$M Rhodamine B dye solution in ethanol.
However, chaotic orbits launched with 
$\chi\gg\chi_c$  can undergo multiple (more than $100$) reflections 
before they reach $\chi_{c}$. Therefore, the leakage loss of a CWGM is much 
less than the round-trip gain, and consequently, lasing can occur 
with feedback provided by chaotic modes.  However, the non-lasing 
four-bounce stable diamond orbits {\it displace} 
the laser emission location of the CWGMs [see Fig. 
1(a), (c)].   Regular WGMs with $\chi\approx  90^\circ$ 
still exist for $L_{z}\approx 0$ 
and lase for this deformation; however their 
leakage rate is much smaller than the absorption rate.   The 
fact that the north and south pole is dark indicates the emission 
from such regular WGM is at least two orders of magnitude weaker 
than that from dynamically eclipsed CWGM for $L_{z}\approx 0$.

\begin{figure}
\hspace*{1.cm}
\epsfig{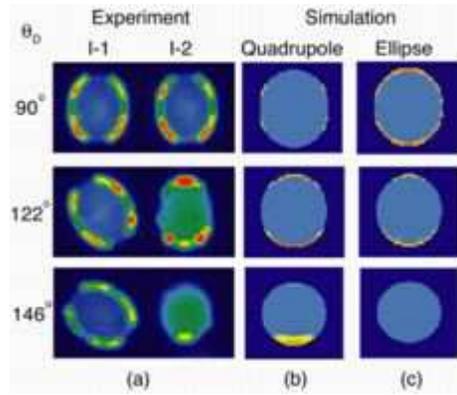}
\caption{(a) Experimental lasing images (I-1 and I-2) of broadly-pumped
(beam diameter: $\!>100\,\mu$m) prolate droplets ($b/a = 1.29$) are
shown for $\theta_D = 90^{\circ}$, $122^{\circ}$, and $146^{\circ}$ (pump
laser was vertically polarized and f/16 lens was used). Corresponding
calculated lasing images (I-2) at $\theta_D = 90^{\circ}$, 
$120^{\circ}$, and
$150^{\circ}$ are shown for quadrupole (b) and elliptic (c) deformations.
In the simulation, equally mixed TE and TM polarizations were used to
calculate Fresnel coefficients for the emerging laser rays.}
\end{figure}
Figure 3 shows additional results consistent with 
this interpretation of dynamical eclipsing of CWGMs; in this case
an unfocused 
beam was used to excite all the possible $L_{z}$-modes.   
For the experimental images at $\theta_{D} = 90^{\circ}$, 
the bright regions along the droplet 
rim are due to a superposition of laser emission from precessing 
modes at different $L_{z}$. 
The abrupt disappearance of emission at the rim 
near the two poles indicates the absence\cite{prl}
of such regular precessing modes with $\chi\approx\chi_{c}$ near 
$L_{z}= 0$.  A very different emission pattern is seen 
in I-2 at tilt angles of $\theta_{D} = 122^{\circ}$ and 
$146^{\circ}$; here we see substantial 
emission in the plane bisecting the poles of the droplet along 
the imaging direction; such emission can only come from $L_{z}\approx 
0$ orbits.  Moreover the fact that the $L_{z}\approx 0$ emission is 
not seen at $\theta_{D}= 90^{\circ}$ insures that the emission does not come 
from the poles (the points of highest curvature), but displaced 
from the two poles in a manner consistent with the dynamical 
eclipsing scenario.

Ray simulations of refractive emission for the unfocused 
pumping case were performed for the quadrupole shape [Fig. 3(b)] 
and ellipsoid [Fig. 3(c)].   Uniformly distributed rays in 
$\theta$  and $L_{z}$ were launched with initial 
$\chi$ greater than $\chi_{c}$.   The lasing condition was imposed in the ray 
simulations by including only orbits which have pathlength (before 
refractive escape) greater than $1/g$, where the gain is estimated
to be $g\approx 45$ cm$^{-1}$.  Ray escape was determined by Snell's law and 
the relevant Fresnel coefficients at each reflection. Note that 
in the quadrupole simulations, there is strong emission near the bottom 
center of the image at $\theta_{D}= 146^\circ$ 
and no emission at the poles of the image at $\theta_{D} = 90^\circ$,
in good qualitative agreement with the experiments\cite{tunneling}. 
In contrast, the ellipsoid simulation shows that most of the 
refractive emission occurs at $\theta_{D}= 90^{\circ}$, 
and essentially no emission is found 
at $\theta_{D}= 146^\circ$ in distinct contrast with our experimental 
results.   This result provides crucial support for our 
interpretation of the images as arising from dynamical eclipsing. 
The ray motion in the ellipsoid is completely regular, with no 
chaotic regions and no four-bounce stable islands [see Fig. 
1(b)]; hence dynamical eclipsing does not occur and one finds 
only tangent emission from the poles [Fig. 1(b), inset].

\begin{figure}
\epsfig{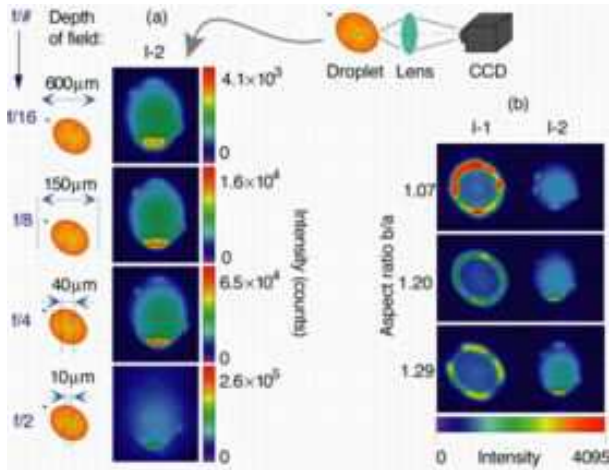}
\caption{(a) Images of broadly-pumped lasing prolate droplets (
$b/a = 1.29$)
at $\theta_D = 142^{\circ}$ with various f-numbers (f/16, f/8, f/4/, and f/2).
Because the solid angle of light acceptance is proportional to 
$\approx 1/(f/\#)^2$,
the color intensity scale for each f/\# is adjusted accordingly. The depth
of field [$\propto(f/\#)^2$] for each 
f/\# is shown relative to the droplet size.
The imaging direction is on the right of the droplet. (b) Lasing images of
broadly-pumped prolate droplets with various aspect ratios. Dynamical
eclipsing is not present for small deformation ($b/a = 1.07$).}
\end{figure}
To further confirm that the emission seen at large 
tilt angles is non-tangent and hence refractive as we expect 
from CWGMs, we have varied the depth of field, $d_{f/\#}$, and the 
acceptance angle (f-number) of the imaging 
system.  In Fig 4(a), images are shown with various $d_{f/\#}$'s 
for a lasing prolate droplets with $b/a = 1.29$ 
at $\theta_{D}= 142^\circ$. The f-number of the camera lens was changed from 
f/2 ($d_{f/2}= 10$ mm, 14\% of the droplet size along the imaging axis) 
to f/16 ($d_{f/16} = 600 \mu$m, covering the whole droplet). While 
the bright emission is sharply imaged at the bottom center of 
the image with f/16, f/8, and f/4, the corresponding emission 
is blurred with f/2.   From this we conclude that this laser 
emission is emitted from outside of the region defined by 
$d_{f/2}$.   If such emission propagates parallel to the 
imaging axis, the emitted rays are non-tangent. For the f/16 image, 
the emission direction is restricted to within $2^\circ$ 
from the imaging axis. 
Laser emission from the displaced 
location implied by the $d_{f/\#}$ analysis 
above is non-tangent by as much as $10^\circ$ - 
$30^\circ$.   Therefore, dynamically eclipsed rays from CWGM 
are non-tangent up to $30^\circ$, which indicates refraction is 
the major escape mechanism.

Further confirmation comes from the aspect-ratio 
dependence of the patterns [Fig. 4(b)].  For moderate deformations 
the 
islands become too small to generate 
dynamical eclipsing, and the expected emission from the polar orbits 
is tangential to the poles as in the case for the ellipse 
[see the inset of Fig. 1(b)]. In Fig. 4(b), lasing images 
at $\theta_{D}= 142^{\circ}$ are shown for various $b/a$; 
$1.07$, $1.20$, and $1.29$.   For $b/a= 1.07$, the entire rim of the 
droplet up to the 
poles is bright in I-1, while no laser emission is recorded in 
I-2 [note the similarity to the ellipsoid simulations in Fig. 
3(c)].  Thus we find a range of intermediate deformations 
with highly anisotropic emission, but no dynamical eclipsing, which 
is then turned on by increasing $b/a$.

In summary, laser emission from a quadrupole-deformed 
spherical microcavity is observed to be partially non-tangent 
and displaced from the highest curvature regions. This provides 
for the first time conclusive evidence of lasing supported by 
chaotic whispering gallery 
modes. These modes are displaced from the high curvature points 
by stable high-loss modes, an effect known as dynamical eclipsing. 
The unique angular dependence of the lasing images is in good 
agreement with simulated images of lasing quadrupolar 
droplets with chaotic ray dynamics.

We would like to acknowledge the partial support 
of NSF (Grant PHY-9612200).      

%******************************************************************

%******************************* Fig. 1 ************************************


\begin{references}
\bibitem{tunneling}
Because the ray simulations do not include tunneling, they cannot
account for the bright rims in the experimental images at $\theta_{D}= 
90^{\circ}$.
\end{references}
\end{document}